\documentstyle[11pt]{article}
\title{Classical Mechanics of Relativistic Particle with Colour}
\author{A. Duviryak\thanks{E--mail: duviryak@icmp.lviv.ua}
\vspace{1ex}\\
{\small\em Institute for Condensed Matter Physics}\\
{\small\em of Ukrainian National Academy of Sciences},\\
{\small\em 1 Svientsitskii Street, UA-290011, Lviv, Ukraine}}
\date{}

\newcommand{\g}{{\hspace{0.1em}{\sl g}\hspace{0.1em}}}
\newcommand{\h}{{\hspace{0.1em}{\sl h}\hspace{0.1em}}}
\newcommand{\p}{{\sl p}}
\newcommand{\G}{{\sl G}}
\newcommand{\Pb}{{\sl P}}
\newcommand{\U}{{\sl U}}
\newcommand{\C}[1]{{\hspace{0.1em}{\sf #1}\hspace{0.1em}}}
\newcommand{\B}[1]{{\hspace{0.1em}{\bf #1}\hspace{0.1em}}}
\newcommand{\Bo}{\mbox{\boldmath$\omega$}}

\newcommand{\Bd}{\mbox{\boldmath$\delta$}}
\newcommand{\ad}{{\rm ad}}
\newcommand{\Ad}{{\rm Ad}}
\newcommand{\IM}{{\rm I}\hspace{-0.18em} {\rm I}\hspace{-0.31em}{\rm M}}
\newcommand{\IR}{{\rm I}\hspace{-0.18em}{\rm R}}

\begin{document}
\maketitle \vspace{-8ex}
\begin{abstract}
Classical description of relativistic pointlike particle with intrinsic
degrees of freedom such as isospin or colour is proposed. It is based on
the Lagrangian of general form defined on the tangent bundle over a principal
fibre bundle. It is shown that the dynamics splits into the external
dynamics which describes the interaction of particle with gauge field in
terms of Wong equations, and the internal dynamics which results in a spatial
motion of particle via integrals of motion only. A relevant Hamiltonian
description is built too.
\vspace{0.5em}
\end{abstract}


\section{Introduction}

    Wong equations of motion of classical pointlike relativistic
particle with isospin or colour \cite{1} permit various Lagrangian and
Hamiltonian formulations. Two of them \cite{4,5} known to the author are
based on geometric notions brought from gauge field theories. Namely, the
configuration space of particle is a principal fibre bundle with
the structure group being the gauge group; the interaction of particle with
an external gauge field is introduced via the connection on this bundle.
The only difference between two approaches consists in the choice of
Lagrangian function. That proposed in \cite{4} is linear in gauge
potentials (as in classical electrodynamics) while the nonlinear
Lagrangian \cite{5,4} arises naturally within the Kaluza--Klein theory.
Nevertheless, both of them lead to the same Wong equations.

    In the present paper, starting from the mentioned above geometrical
treatment of the kinematics of relativistic particle with isospin or colour
(Section 2), we construct in Section 3 the Lagrangian of a fairy general
form. In particular cases this Lagrangian reduces to those of Refs.
\cite{4,5}. The generalization is not trivial since it leads to
two types of dynamics, according to what variables are used
for the description of intrinsic degrees of freedom. The external dynamics is
described by the Wong equations which include gauge potentials, but the form
of which is indifferent to the choice of Lagrangian. On the contrary, the
internal dynamics is governed by the particular choice of Lagrangian function
while gauge potentials completely fall out of this dynamics. This fact
becomes more transparent within the framework of the appropriate Hamiltonian
formalism developed in Section 4. In Section 5 we sum up our
results and discuss the perspectives of quantization.


\section{Kinematics on a principal fibre bundle}

    Following Refs. \cite{4,5} we take for a configuration space of
relativistic particle with isospin or colour the principal fibre bundle $\Pb$
over Minkowski space $\IM$ with structure group $\G$ and projection $\pi:
\Pb \to \IM$
(see Ref. \cite{6} for these notions). A particle trajectory $\gamma: \IR
\to \Pb$; $\tau\mapsto \p(\tau) \in \Pb$ is parameterized by evolution
parameter $\tau$. A state of particle is determined by $(\p, \dot\p)
\in {\rm T}P$.

    In this paper we are not interested in the global structure of $\Pb$,
and use its local coordinatization: $\Pb\ni\p = (x,\g) = (x^\mu, g^i),\
\mu = \overline{0,3},\ i = \overline{1,\dim\G}$, where $x = \pi(\p) \in
\U\subset\IM$ ($\U$ is an open subset of $\IM$), $\g = \varphi(\p) \in\G$,
and $\varphi: \Pb\to\G$ defines the choice of gauge. Respectively, $\dot\p =
(\dot x, \dot\g)$, where $\dot x \in {\rm T}_x\IM$ and $\dot\g \in
{\rm T}_{\g}\G$. We call $(x,\dot x)$ and $(\g,\dot\g)$ the space and the
intrinsic (local) variables of particle respectively.

    Let the principal fibre bundle $\Pb$ be endowed with a connection
defined by 1-form $\Bo$ on $\Pb$ which takes values in Lie algebra $\cal G$
of $\G$. Locally it can be represented as follows \cite{6}:
%
\begin{equation} \label{2.1}
\Bo = \Ad_{\g^{-1}}\pi^*(\B A_\mu(x)dx^\mu) + \g^{-1}d\g \equiv
\g^{-1}\left(\pi^*(\B A_\mu(x)dx^\mu)\right)\g + \g^{-1}d\g,
\end{equation}
where $\pi^*$ is the pull back mapping onto $\Pb$, Ad denotes the
adjoint representation of $\G$ in $\cal G$, and $\cal G$-valued functions
$\B A_\mu(x)$ are gauge potentials. Under a right action of $\G$ defined
in $\Pb$ by
%
\begin{equation} \label{2.3}
R_\h: \p\mapsto \p'\equiv R_\h(x,\g) = (x,\g\!\h),
\qquad \h \in \G,
\end{equation}
the connection form transforms via a pull back mapping $R_\h^*$ and
is equivariant, i.e., $R_\h^*\Bo = \Ad_{\h^{-1}}\Bo$,
$\h \in \G$.

    A gauge transformation arises in a geometrical treatment as a
bundle automorphism defined by
%
\begin{equation} \label{2.4}
\Phi_{\h(x)}: \p\mapsto \p'\equiv \Phi_{\h(x)}(x,\g)
= (x,\h(x)\g),
\qquad \h(x)\in\G.
\end{equation}
It induces the transformation of the connection form defined by the inverse
of the pull back mapping (actually, a push forward mapping),
$\Bo\to\Bo'=[\Phi_{\h^{-1}(x)}]^*\Bo$, so that the value of the connection
form on each vector field is gauge-invariant by definition. The transformed
form $\Bo'$ is also expressed by eqs. (\ref{2.1}), but with
new potentials
%
\begin{equation} \label{2.5}
\B A'_\mu(x) = \h(x)\B A_\mu(x)\h^{-1}(x) +
\h(x)\partial_\mu\h^{-1}(x).
\end{equation}

    The Minkowski metrics
$\eta\equiv \eta_{\mu\nu}dx^\mu\otimes dx^\nu$,
$\|\eta_{\mu\nu}\|={\rm diag}\,(+,-,-,-)$
is defined on base space $\IM$. It is invariant under the Poincar\'e group
acting in $\IM$. Being pulled back by $\pi^*$ onto $\Pb$ it becomes also
right- and gauge-invariant, but appears degenerate.

    Here we suppose that the Lie algebra $\cal G$ of structure
group $\G$ is endowed with non-degenerate Ad-invariant metrics
$\langle~\cdot~,~\cdot~\rangle$. The example is the Killing-Cartan metrics in
the case of semi-simple group. In terms of this metrics, the connection form,
and the Minkowski metrics one can construct a nondegenerate metrics on the
bundle $\Pb$ \cite{5},
%
\begin{equation} \label{2.7}
\Xi = \pi^*\eta - a^2\langle\Bo,\Bo\rangle
\end{equation}
($a$ is a constant), which is right- and gauge-invariant but not Poincar\'e-%
invariant
(the latter is broken by $\Bo$). In the case of bundle over a curved base
space the Minkowski metrics on the right-hand side (r.h.s.) of eq. (\ref{2.7})
is replaced by the Riemanian one. In this form the metrics $\Xi$ arises in
the Kaluza-Klein theory \cite{7} which allows to unify the description of
gravitational and Yang-Mills fields.


\section{Lagrangian dynamics of particle with isospin or colour}

    The dynamical description of the relativistic particle with isospin
or colour should, at least, satisfy the following conditions:\vspace{-1.2ex}
\begin{description}
\item[i)] gauge invariance;\vspace{-1.8ex}
\item[ii)] invariance under an arbitrary change of evolution parameter;
\vspace{-1.8ex}
\item[iii)] Poincar\'e invariance provided gauge potentials vanish.
\vspace{-1.2ex}
\end{description}

    These demands can be embodied in the action
$I = \int d\tau L(\p,\dot\p)$
%
%
with the following Lagrangian
%
\begin{equation} \label{3.2}
L = |\dot x|F(\B w),
\end{equation}
where $\B w\equiv\Bo(\dot\p)/|\dot x|$,~~$|\dot x|\equiv\sqrt{\eta_{\mu\nu}
\dot x^\mu\dot x^\nu} = \sqrt{\pi^*\eta(\dot\p,\dot\p)}$, and $F: \cal
G\to\IR$ is an arbitrary function. We note that the quantities
$|\dot x|$, $\Bo(\dot\p)$, and thus the variable $\B w$ and the Lagrangian
(\ref{3.2}) are gauge-invariant.

    In order to calculate the variation $\delta\!I$ of the action
$I$ it is convenient to use, instead of intrinsic velocity $\dot
\g$ and variation $\delta\!\g$, the following variables: $\B v\equiv
\dot\g\g^{-1}$ and $\Bd\!\g\equiv\delta\!\g\g^{-1}$. They take
values in Lie algebra $\cal G$ of group $\G$. Then the argument $\B w$ of
$F$ can be presented in the form
%
\begin{equation} \label{3.3}
\B w = \Ad_{\g^{-1}}(\B v + \B A_\mu\dot x^\mu)/|\dot x|.
\end{equation}

    Using the formal technique (see Ref. \cite{8} for rigorous
substantiation)
%
\begin{eqnarray} \label{3.4}
\delta\!\g^{-1}\!\!&=& -\g^{-1}\delta\!\g\g^{-1}, \nonumber \\
\delta\!\B v&=&\delta(\dot\g\g^{-1})\ =\
\delta\!\dot\g\g^{-1} + \dot\g\delta\!\g^{-1}
\ =\ \frac{d}{d\tau}\Bd\!\g - [\B v,\Bd\!\g],
\nonumber \\
\delta(\Ad_{\g^{-1}}\B V)&=&\delta(\g^{-1}\B V\g)\ =\
\Ad_{\g^{-1}}(\delta\!\B V + [\B V,\Bd\!\g])
\end{eqnarray}
etc., where $[~\cdot~,~\cdot~]$ are Lie brackets in $\cal G$ and $\B V$ is
an arbitrary $\cal G$-valued quantity, we obtain the following Euler-Lagrange
equations:
%
\begin{eqnarray}
&\dot p_\mu = \C q\cdot(\partial_\mu\B A_\nu)\dot x^\nu,&\label{3.5}\\
&\dot\C q\ =\ \ad_{\B A\cdot\dot x}^*\C q&\label{3.6}
\end{eqnarray}
with
%
\begin{eqnarray}
&\displaystyle{p_\mu\equiv \frac{\partial L}{\partial\dot x^\mu}\ =\
\left(F - \frac{\partial F}{\partial\B w}
\cdot\B w\right)\frac{\dot x_\mu}{|\dot x|} + \C q\cdot\B A_\mu,}&
\label{3.7}\\
&\displaystyle{\C q\equiv\frac{\partial L}{\partial\B v}\ =\
\Ad_{\g^{-1}}^*\frac{\partial F}{\partial\B w},}&\label{3.8}
\end{eqnarray}
where $p_\mu$ are spatial momentum variables, and $\C q$
is an intrinsic momentum-type variable which takes values in co-algebra
$\cal G^*$.
The linear operators $\ad^*$ and $\Ad^*$ define co-adjoint representations
of $\cal G$ and $\G$ respectively, dot ``~$\cdot$~'' denotes a contraction.

    First of all we show that the function
%
\begin{equation} \label{3.10}
M(\B w)\equiv F - \frac{\partial F}{\partial\B w}\cdot\B w
\end{equation}
is an integral of motion. For this purpose let us introduce the following
$\cal G^*$-valued variable:
%
\begin{equation} \label{3.11}
\C s\equiv\frac{\partial F}{\partial\B w}\ =\  \Ad_\g^*\C q.
\end{equation}
In contrast to $\C q$, it is gauge-invariant.
Taking into account the equations (\ref{3.11}) and
(\ref{3.6}) we obtain after a bit calculation the equation:
%
\begin{equation} \label{3.14}
\dot\C s\ =\ \ad_{|\dot x|\B w}^*\C s.
\end{equation}
Then
$\dot M = -\dot\C s\cdot\B w = -\C s\cdot[\B w,\B w]|\dot x|\equiv 0$
q.e.d.

    Using this fact and (\ref{3.6}) in (\ref{3.5}) yields the
equations of spatial motion
%
\begin{equation} \label{3.16}
M\frac{d}{d\tau}\frac{\dot x_\mu}{|\dot x|} =
\C q\cdot\B F_{\mu\nu}\dot x^\nu,
\end{equation}
where
%
$\B F_{\mu\nu} \equiv \partial_\mu\B A_\nu - \partial_\nu\B A_\mu +
[\B A_\mu,\B A_\nu].$
%
Equations (\ref{3.16}) together with the equation (\ref{3.6}) or (\ref{3.14})
of intrinsic motion determine the particle dynamics on a principal fibre
bundle.

    Now we suppose the existence in Lie algebra $\cal G$ of non-degenerate
metrics $\langle~\cdot~,~\cdot~\rangle$. It allows to identify $\cal G^*$ and
$\cal G$. In particular, for co-vector $\C q\in\cal G^*$ we introduce
the corresponding vector $\B q\in\cal G$, such that
$\C q = \langle\B q,~\cdot~\rangle$.
Then the equations of motion (\ref{3.16}) and (\ref{3.6}) take
the form
%
\begin{eqnarray}
&\displaystyle{M\frac{d}{d\tau}\frac{\dot x_\mu}{|\dot x|} =
\langle\B q,\B F_{\mu\nu}\rangle\dot x^\nu}, & \label{3.19} \\
&\dot\B q\ =\ [\B q,\B A_\mu]\dot x^\mu.& \label{3.20}
\end{eqnarray}
Besides, if this metrics is Ad-invariant, the quantity $\langle\B q,\B q
\rangle$ is an integral of motion.

    At this stage we have obtained the well-known Wong equations
(\ref{3.19})--(\ref{3.20}) which describe the dynamics of a relativistic
particle with mass $M$ and isospin or colour $\B q$. Despite that we started
with the Lagrangian (\ref{3.2}) of a fairy general form, this arbitrariness
is obscured in the Wong equations. The reason resides in definitions of
$M$ and $\B q$ which are, in general, complicated functions on
${\rm T}\Pb$. This feature is better understood by analyzing
equations (\ref{3.16}) and (\ref{3.6}) which are very similar to the
Wong equations but do not involve the metrics in $\cal G$.

    In general, the set of eqs. (\ref{3.16}) and (\ref{3.6}) is of the
second order with respect to configuration variables $x$ and $\g$. In this
regard it is quite equivalent to the set of eqs. (\ref{3.16}) and (\ref{3.14}).
On the other hand, the equations (\ref{3.16}) and (\ref{3.6}) form a self-%
contained set in terms of variables $x$ and $\C q$. They involve explicitly
potentials of external gauge field, but their form is indifferent to a choice
of Lagrangian.

    The equation (\ref{3.14}) is the closed first-order equation with
respect to $\B w$ or, if eq. (\ref{3.11}) is invertible, with respect to
$\C s$ (the quantity $|\dot x|$ is not essential because of
a parametric invariance of dynamics; we can put, for instance, $|\dot x|=1$).
In contrast to the set (\ref{3.16}), (\ref{3.6}), the equation (\ref{3.14})
is determined by a choice of the Lagrangian, but, in terms of $\B w$ or
$\C s$, it does not include gauge potentials.

    Hence, the dynamics of isospin particle splits into the {\em
external dynamics} described by the equations (\ref{3.16}), (\ref{3.6})
in terms of variables $x$ and $\C q$, and
the {\em internal dynamics} determined by the equation (\ref{3.14}) in
terms of $\B w$ or $\C s$. The only coupling of these
realizations of dynamics is provided via integrals of motion, namely, the
particle mass $M$ and (if Ad-invariant metrics is involved) the isospin
module $|\B q|\equiv\sqrt{\langle\B q,\B q\rangle} = |\B s|$.

    In the following examples we show that the general description of
isospin particle includes, as particular cases, results known in the
literature. Besides, we demonstrate some new features concerning the
internal dynamics.


\par
    {\bf 1. Linear Lagrangian. Electrodynamics.}
    The simplest choice of Lagrangian (\ref{3.2}) leading to non-trivial
intrinsic dynamics corresponds to the following function $F(\B w)$:
%
\begin{equation} \label{4.1}
F(\B w) = m + \C k\cdot\B w,
\end{equation}
where $m\in\IR$ and $\C k\in\cal G^*$ are constants. Up to notation it
coincides with the Lagrangian proposed by Balachandran et al. in Ref.
\cite{4}. In this case the isospin $\C q=\Ad_{\g^{-1}}^*\C k$ is purely
configuration variable (it does not depend on velocities) and the equation
(\ref{3.6}) is truly the first-order Euler-Lagrange equation. Besides,
$M$ and $\C s$ are constants, i.e., $M=m$, ~$\C s=\C k$, thus the internal
dynamics completely degenerates.

    The Lagrangian (\ref{3.2}), (\ref{4.1}) is linear with respect to
gauge potentials $\B A_\mu$. In the case of one-parametric gauge group $U(1)$
it reduces to that of electrodynamics. Indeed, in this case we have
$\g= \exp({\rm i}\theta)$, $\B v={\rm i}\dot\theta$, $\B A_\mu={\rm i}A_\mu$.
Choosing $\C k=-{\rm i}e$, where $e$ is the charge of electron, one can
present the Lagrangian in the form:
%
\begin{equation} \label{4.3}
L = m|\dot x| + eA_\mu\dot x^\mu + e\dot\theta.
\end{equation}
The third term on r.h.s. of (\ref{4.3}) is a total derivative and thus it can
be omitted. Hence, intrinsic variables disappear in this Lagrangian, and the
latter takes the standard form.

    The similar situation occurs when considering an arbitrary Abelian
gauge group.

    {\bf 2. Right-invariant Lagrangian. Kaluza-Klein theory.}
    The following choice of the function $F(\B w)$:
%
\begin{equation} \label{4.6}
F(\B w) = f(|\B w|),\qquad
|\B w|\equiv\sqrt{\langle\B w,\B w\rangle} =
|\B v+\B A_\mu\dot x^\mu|/|\dot x|,
\end{equation}
where $f:\IR\to\IR$ is an arbitrary function of $|\B w|$, corresponds to
Lagrangian which is invariant under the right action of $\G$. Following the
Noether theorem there exist corresponding
integrals of motion. In the present case they form $\cal G^*$-valued
co-vector $\C s$ defined by eq. (\ref{3.11}) or, equivalently, $\cal G$-%
valued vector
%
\begin{equation} \label{4.8}
\B s=|\B w|^{-1}f'(|\B w|)\B w,
\end{equation}
where $f'(|\B w|)\equiv d\,f/d\,|\B w|$. We note that the integrals of motion
$\B s$ and $M(\B w)$ are not independent. Indeed, if $f'(|\B w|)$ is not
constant, the mass $M$ can be presented as a function of $|\B s|=|\B q|$.
Otherwise both these quantities are constants. Thus the mass $M$ and the
isospin module $|\B q|$ are completely determined by the external dynamics.

    The right-invariant Lagrangian of special kind arises naturally
in the framework of Kaluza-Klein theory \cite{7}. It has the following
form \cite{5,4}:
%
\begin{equation} \label{4.9}
L = m\sqrt{\Xi(\dot\p,\dot\p)},
\end{equation}
where the metrics $\Xi$ on a principle bundle $\Pb$ is introduced by eq.
(\ref{2.7}). In our notations this Lagrangian corresponds to the choice:
%
\begin{equation} \label{4.10}
f(|\B w|) = m\sqrt{1-a^2|\B w|^2}.
\end{equation}
This function determines the following relation between $M$ and $|\B q|$:
%
\begin{equation} \label{4.11}
M(|\B q|) = \sqrt{m^2+|\B q|^2/a^2}.
\end{equation}

    {\bf 3. Isospin top.}
    In the above two examples the internal dynamics does not affect the
external dynamics. Here we consider a contrary example. Let
%
\begin{equation} \label{4.12}
F(\B w) = f(\nu),\qquad\nu\equiv\sqrt{\langle\B w,T\B w\rangle},
\end{equation}
where $T$ is a self-adjoint (in the metrics $\langle~\cdot~,~\cdot~\rangle$)
linear operator. In this case we have
%
\begin{equation} \label{4.13}
\B s = \nu^{-1}f'(\nu)T\B w, \qquad
M = f(\nu) - \nu f'(\nu).
\end{equation}
If the function $f(\nu)$ is not linear, the quantity $\nu$ turns out to be
an integral of motion which is independent of $|\B q|$. Then using the
parameterization $|\dot x|=1$ one can reduce eq.(\ref{3.14}) to the following
equation of internal motion:
%
\begin{equation} \label{4.15}
T\dot\B w = [T\B w,\B w].
\end{equation}
This is nothing but the compact form of Euler equations (i.e., the equations
of motion of a free top) generalized to the case of arbitrary group
\cite{9}. A solution of this equation is necessary for evaluation of the
observable mass of particle.

    The relation between the external dynamics and the internal one
becomes more transparent within the Hamiltonian formalism which we consider
in the next section.


\section{Transition to Hamiltonian description}

    The Lagrangian description on the configuration space $\Pb$ enables
a natural transition to the Hamiltonian formalism with constraints \cite{10}
on the cotangent bundle ${\rm T}^*\Pb$ over $\Pb$. Locally, ${\rm T}^*\Pb
\simeq{\rm T}^*\U\times{\rm T}^*\G$ and, in turn, ${\rm T}^*\G\simeq
\G\times{\cal G}^*$. The latter isomorphism is established by right or
left action of group $\G$ on ${\rm T}^*\G$ (see, for instance, \cite{11}).
It is implicitly meant in our notation. Namely, we coordinatize
${\rm T}^*\G$ by variables $(\g,\C q)$ or $(\g,\C s)$.

    Let us introduce basis vectors $\B e_i\in{\cal G}$ satisfying
the Lie-bracket relations $[\B e_i,\B e_j] = c_{ij}^k\B e_k$,
where $c_{ij}^k$ are the structure constant of $\G$, and basis co-vectors
$\C e^i\in{\cal G}^*$ such that $\C e^j\cdot\B e_i=\delta^j_i$. Then the
standard symplectic structure on the cotangent bundle ${\rm T}^*\Pb$ over
the manifold $\Pb$ can be expressed in terms of
local coordinates $x^\mu$, $p_\nu$, $g^i$ and $q_j\equiv\C q\cdot\B e_j$
by the following Poisson-bracket relations:
%
\begin{equation} \label{5.2}
\{x^\mu,p_\nu\}=\delta^\mu_\nu,\qquad
\{q_i,q_j\} = c_{ij}^kq_k,\qquad
\{g^i,q_j\} = \zeta^i_j(\g)
\end{equation}
(other brackets are equal to zero),
where $\zeta_i^j(\g)$ are the components of right-invariant vector fields
on $\G$. Equivalently, we can use variables
$s_j\equiv\C s\cdot\B e_j$ instead of $q_j$. Then the Poisson-bracket
relations take the form:
%
\begin{equation} \label{5.4}
\{x^\mu,p_\nu\}=\delta^\mu_\nu,\qquad
\{s_i,s_j\} = -c_{ij}^ks_k,\qquad
\{g^i,s_j\} = \xi^i_j(\g),
\end{equation}
where $\xi_i^j(\g)$ are the components of left-invariant vector fields
on $\G$.

Once the Poisson brackets are defined, we can use in calculations both sets
of variables. In particular, taking into account the relation
$\C s=\Ad_\g^*\C q$ we obtain:
%
\begin{equation} \label{5.6}
\{q_i,s_j\} = 0.
\end{equation}

    The transition from the Lagrangian description to the Hamiltonian one
lies through the Legendre transformation defined by eqs. (\ref{3.7}) and
(\ref{3.8}) or (\ref{3.11}). It is degenerate and leads to the vanishing
canonical Hamiltonian, due to parametrical invariance. Instead, the dynamics
is determined by constraints.

    In order to obtain constraints explicitly let us consider the
relations (\ref{3.10}) and (\ref{3.11}). They present, in fact, the
Legendre mapping $\B w\mapsto\C s$ and thus allow to consider the mass
$M$ as a function of $\C s$ only. Then eq. (\ref{3.7}) reduces to
%
\begin{equation} \label{5.7}
\Pi_\mu\equiv p_\mu - \C q\cdot\B A_\mu =
M(\C s)\dot x_\mu/|\dot x|
\end{equation}
which yields immediately the {\em mass-shell constraint}:
%
\begin{equation} \label{5.8}
\phi\equiv\Pi^2 - M^2(\C s) = 0.
\end{equation}

    There are no more constraints if
%
\begin{equation} \label{5.9}
\det\left\|\frac{\partial^2 F(\B w)}{\partial w^i\partial w^j}\right\| \ne 0.
\end{equation}
Otherwise, eq. (\ref{3.11})
leads to additional constraints of the following general structure:
%
\begin{equation} \label{5.10}
\chi_r(\C s) = 0,\qquad r=\overline{1,\kappa}\le\dim\G
\end{equation}
which together with the mass-shell constraint (\ref{5.8}) form the set of
primary constraints. Hence, the Dirac Hamiltonian is
$H_D = \lambda_0\phi + \lambda^r\chi_r$,
where $\lambda_0,\lambda^r$ are Lagrange multipliers. It is evident
from eqs. (\ref{5.8}), (\ref{5.10}) and (\ref{5.2}),
(\ref{5.4}), (\ref{5.6}) that secondary constraints (should they exist)
are of the same general structure as in eq. (\ref{5.10}). At the final stage
of analysis the mass-shell constraint can be considered as the first-class
one. This is provided as soon the mass squared $M^2(\C s)$ is conserved. If
it is not, there exists another integral of motion $\tilde M^2(\C s)$ such
that $\tilde M^2(\C s)|_{\chi=0}=M^2(\C s)$. Then the first-%
class mass-shell constraint has the form (\ref{5.8}) but with the function
$\tilde M^2(\C s)$ instead of $M^2(\C s)$.

    The total set of constraints is gauge-invariant. This follows from
the transformation properties of gauge potentials (\ref{2.5}) and variables:
%
\begin{equation} \label{5.12}
{x^\mu}' = x^\mu,\quad
{p_\mu}' = p_\mu -
\C q\cdot\left(\h^{-1}(x)\partial_\mu\h(x)\right),\quad
{\C q}' = \Ad_{\h^{-1}(x)}^*\C q,\quad {\C s}' = \C s.
\end{equation}
In particular, the variables $\Pi_\mu$
defined in eq. (\ref{5.7}) are gauge-invariant.

    At this stage the splitting of particle dynamics into the external
and internal ones becomes obvious. Indeed, the Hamiltonian equations
%
\begin{equation} \label{5.13}
(\dot x,\dot p,\dot{\C q}) = \{(x,p,\C q), H_D\} \approx
\lambda_0\{(x,p,\C q),\Pi^2\},
\end{equation}
where $\approx$ is Dirac's symbol of weak equality, are closed with respect
to variables $(x,p,\C q)$. They describe the external dynamics and can be
reduced to the equations  (\ref{3.16}) and (\ref{3.6}) by eliminating the
variables $p_\mu$ and the multiplier $\lambda_0$. The equation
%
\begin{equation} \label{5.14}
\dot\C s = \{\C s, H_D\} \approx
-\lambda_0\{\C s,M^2(\C s)\} + \lambda^r\{\C s,\chi_r(\C s)\}
\end{equation}
is closed in terms of $\C s$ and can be reduced to the equation
(\ref{3.14}) of the internal dynamics. We note that the group
variable $\g$ falls out of the equations (\ref{5.13}) and (\ref{5.14})
which is due to the structure of Poisson-bracket relations (\ref{5.2}),
(\ref{5.4}), (\ref{5.6}) and constraints (\ref{5.8}), (\ref{5.10}). Thus
in the present formulation of isospin particle dynamics this variable
can be considered as redundant unobservable quantity.

    The further treatment of Hamiltonian dynamics, i.e., the
classification of constrains as first- and second-class ones etc., demands a
consideration of some specific examples.


\section{Conclusions}

    In this paper we consider the formulation of classical dynamics of
the relativistic
particle in an external Yang-Mills field. We have deduced the Wong equations
from the Lagrangian of rather general form defined on the tangent bundle over
principle fibre bundle. Besides, we have shown that this Lagrangian leads to
some internal particle dynamics. The only
quantities coupling this dynamics with the Wong equations are the mass $M$
and isospin (or colour) module $|\B q|$, the intrinsic characteristics of
particle. In the present description they are integrals of internal motion.

    The physical treatment of internal dynamics should become better
understood within an appropriate quantum-mechanical description. It can be
constructed on the base of Hamiltonian particle dynamics proposed in
Section 4. Here we only suggest some features of such a description.

    Following the procedure of canonical quantization one replaces
dynamical variables $x,p,\C q,\C s$ etc. by operators $\hat x,\hat p,
\hat \C q,\hat \C s$, and Poisson brackets by commutators.
Let us suppose that the classical dynamics is determined by the only
mass-shell constraint (\ref{5.8}). Its quantum analogue
determines physical states of the system. Eigenvalues $q$ and $M$ of
operators $|\hat\B q|^2=|\hat\B s|^2$ and $M^2(\hat\C s)$ which commute
with mass-shell constraint and with one another can be treated as the
isospin (or colour) and the rest mass of particle. In the case of
right-invariant dynamics (as in Kaluza-Klein theory) $M$ is unambiguous
function of $q$. In the general case, the spectrum of $M^2(\hat\C s)$ can
consist of few levels $M_{qn}$ which correspond to the same value of $q$.
Thus it is tempting to relate the quantum number $n$ with a flavour or
generation. Of course, this supposition is by no
means substantiated and needs a following elaboration. It may
suggest a phenomenological quantum
description of relativistic particles with intrinsic degrees of freedom type
of isospin, colour, flavour etc.\\

\subsection*{Acknowledgments}
\par\
    The author would like to thank Prof. V. Tretyak and Drs. R. Matsyuk,
A. Panasyuk, V. Shpytko, and Yu. Yaremko for stimulating discussions of this
work.

\end{document}